\documentclass[prd,nopacs,nokeys,nofootinbib]{revtex4}
\usepackage{amsfonts}
\usepackage{amsmath}
\usepackage{amssymb}
\usepackage{bbm}
\usepackage[utf8]{inputenc}
\usepackage[caption=false]{subfig}
\usepackage{tensor}
\usepackage{slashed}
\usepackage[centertableaux]{ytableau}
\usepackage{hyperref}
\usepackage{natbib}
\usepackage{braket,mleftright}
\usepackage{graphicx}
\usepackage{cleveref}
\usepackage{tikz}
\usepackage{booktabs}
\usepackage{tikz}
\usepackage{amsmath}

\usetikzlibrary{decorations.pathmorphing, decorations.markings, positioning, arrows.meta}

\tikzset{
    fermion/.style={
        draw=black, 
        thick, 
        postaction={decorate},
        decoration={markings, mark=at position 0.55 with {\arrow{Latex[length=2.5mm, width=2.5mm]}}}
    },
    gauge/.style={
        draw=black, 
        thick, 
        decorate, 
        decoration={snake, amplitude=1.5pt, segment length=5pt}
    },
    vertex/.style={
        circle, 
        draw=black, 
        fill=black, 
        inner sep=1.5pt
    }
}

\begin{document}

\title{Fractional epidemics from quantum loops}

\author{Jos\'e de Jes\'us Bernal-Alvarado}
\email{bernal@ugto.mx}
\affiliation{Physics Engineering Department, Universidad de Guanajuato, M\'{e}xico}

\author{David Delepine}
\email{delepine@ugto.mx}
\affiliation{Physics Department, Universidad de Guanajuato, M\'{e}xico}

\date{\today}

\begin{abstract}
Classical compartmental models of epidemiology rely on well-mixed, local interaction approximations that fail to capture the heavy-tailed burst dynamics and long-range spatial correlations observed in real-world outbreaks. While fractional calculus is frequently employed to model these anomalous behaviors, fractional operators are  introduced phenomenologically. In this work, we demonstrate that fractional space-time epidemic dynamics emerge naturally and rigorously from first principles using a non-equilibrium quantum field theory model. By mapping the stochastic contagion process to a gauge-mediated field theory via the Doi-Peliti formalism, we go beyond the static mean-field approximation to compute the full dynamical one-loop vacuum polarization. We prove that integrating out a dynamically fluctuating host vacuum generates anomalous momentum and frequency scaling. Transitioning back to coordinate space, this derives a coupled space-time fractional integro-differential equations, where the non-linear transmission vertex is governed by parabolic Riesz potentials and Riemann-Liouville time derivatives. We show that in the anomalous regime ($\alpha < 2$), local Debye screening is modified, facilitating Lévy flight super-spreading and temporal avalanches. Consequently, the effective reproductive number ($R_{eff}$) ceases to be a scalar, transforming into a spectral dispersion relation bounded strictly by the ultraviolet spatial cutoff.
\end{abstract}

\maketitle
%%%%%%%%%%%%%%%%%%%%%%%
\section{Introduction}
%%%%%%%%%%%%%%%%%%%%%%%
The mathematical modeling of infectious diseases has historically been anchored by the classical Susceptible-Infected-Recovered (SIR) model and its derivatives \cite{kermack1927contribution}.  Formulated as a system of ordinary differential equations (ODEs), the standard SIR model relies implicitly on the Central Limit Theorem and the assumption of a well-mixed, homogeneous population. These assumptions mathematically enforce a  Markovian and local physical reality: transmission is instantaneous, biological residence times follow exponential distributions, and the pathogen propagates through standard Brownian diffusion.

However, empirical data from modern pandemics reveals a  different topology. Real-world transmission is characterized by extreme heterogeneity, driven by super-spreader events \cite{lloyd2005superspreading}, scale-free human mobility networks \cite{brockmann2006scaling}, and complex, non-exponential incubation periods. Under these conditions, the variance of the microscopic transmission events diverges, leading to the breakdown of classical diffusion and the emergence of anomalous scaling \cite{brockmann2006scaling}.

To capture these non-local and super-diffusive behaviors, researchers have increasingly turned to fractional calculus \cite{lu2023anomalous, yang2025fractional}. Standard fractional epidemic models usually inject separate, decoupled operators: the Riemann-Liouville (or Caputo) derivative for time, and the Riesz fractional Laplacian for space. While successful at fitting heavy-tailed epidemiological data, standard approaches  introduce fractional operators phenomenologically as linear transport terms to model the anomalous movement of hosts \cite{yang2025fractional}, while leaving the transmission vertex  local. This ad-hoc insertion lacks a  microscopic foundation \cite{lu2023anomalous}.

In this paper, we propose a unifying theoretical model by mapping the stochastic microscopic interactions of an epidemic onto a non-equilibrium quantum field theory. Utilizing the Doi-Peliti formalism \cite{doi1976second,peliti1985path,tauber2014critical}, we model the contagion not as an instantaneous point-contact, but as an interaction mediated by a dynamic pathogen gauge field,a mechanism we recently formalized as gauge-mediated contagion \cite{bernalalvarado2026gaugemediated}. In first approximation, the susceptible field is usually treated as a static, homogeneous background condensate. We demonstrate that by allowing the susceptible background to behave as a dynamical field with its own stochastic fluctuations, the one-loop vacuum polarization can generate anomalous momentum scaling.

We prove that fractional integro-differential equations are not mere phenomenological tools, but the exact macroscopic consequence of integrating out this dynamically fluctuating host vacuum. By deriving the spectral effective reproductive number, $R_{eff}(k, \omega)$, we reveal how fractional kinematics systematically destroy classical local herd immunity, forcing macroscopic transmission to become hypersensitive to microscopic spatial constraints.

%\section{The Epidemic Gauge Theory Model}

\section{Dynamical Vacuum Polarization (Beyond the Static Limit)}

\begin{figure}[htbp]
    \centering
    \begin{tikzpicture}
        % Define custom styles for Feynman diagrams
        \tikzset{
            % Host population (solid line with arrow)
            fermion/.style={
                draw=black, 
                thick, 
                postaction={decorate},
                decoration={markings, mark=at position 0.55 with {\arrow{Latex[length=2.5mm, width=2.5mm]}}}
            },
            % Pathogen gauge field (wavy line)
            gauge/.style={
                draw=black, 
                thick, 
                decorate, 
                decoration={snake, amplitude=1.5pt, segment length=5pt}
            },
            % Interaction vertex (black dot)
            vertex/.style={
                circle, 
                draw=black, 
                fill=black, 
                inner sep=1.5pt
            }
        }

        % Define coordinates
        \coordinate (Phi_in) at (-2.5, 0);
        \coordinate (V1) at (0, 0);    % Left vertex (beta_0)
        \coordinate (V2) at (3.5, 0);  % Right vertex (g)
        \coordinate (Phi_out) at (6, 0);

        % External pathogen lines
        \draw[gauge] (Phi_in) -- node[above] {$\varphi(k, \omega)$} (V1);
        \draw[gauge] (V2) -- node[above] {$\varphi(k, \omega)$} (Phi_out);

        % Virtual Host Loop
        % Upper arc (Infected propagator): flows left to right
        \draw[fermion] (V1) arc[start angle=180, end angle=0, radius=1.75cm] 
            node[midway, above=0.15cm] {$I(k-q, \omega-\Omega)$};
            
        % Lower arc (Susceptible propagator): flows right to left
        \draw[fermion] (V2) arc[start angle=0, end angle=-180, radius=1.75cm] 
            node[midway, below=0.15cm] {$S(q, \Omega)$};

        % Draw the vertices and add labels
        \node[vertex, label=right:{$\beta_0$}] at (V1) {};
        \node[vertex, label=left:{$g$}] at (V2) {};

    \end{tikzpicture}
    \caption{One-loop Feynman diagram for the dynamical vacuum polarization $\Pi(k,\omega)$ (pathogen self-energy). The external pathogen field $\varphi$, carrying momentum $k$ and frequency $\omega$, fluctuates into a virtual pair composed of a Susceptible host $S$ and an Infected host $I$. The loop represents the integral over all possible internal momenta $q$ and frequencies $\Omega$ of the fluctuating host vacuum. The left vertex corresponds to the transmission absorption event ($\beta_0$), while the right vertex corresponds to the subsequent emission ($g$).}
    \label{fig:vacuum_polarization}
\end{figure}
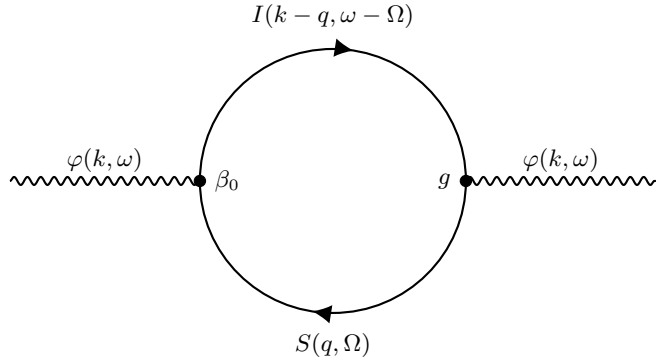
We shall use the model of gauge-mediated contagion proposed in ref.\cite{bernalalvarado2026gaugemediated}.
To capture the non-local, super-diffusive behaviors inherent in complex epidemic dynamics, the theoretical model must go beyond the mean-field approximation of the susceptible population \cite{cardy1985epidemic}. In first approximation,  the susceptible field is usually  treated as a static, homogeneous background condensate. By allowing the susceptible background to behave as a  dynamical field with its own stochastic fluctuations, the one-loop vacuum polarization can  generate anomalous momentum scaling \cite{metzler2000random}.

The vacuum polarization, or self-energy $\Pi(k,\omega)$, represents how the host population ``responds'' to the presence of the pathogen field, effectively altering the transmission environment \cite{janssen2005field}. It is calculated by evaluating the one-loop Feynman diagram, which involves integrating over the internal momentum $q$ and frequency $\Omega$ of the virtual host fields:

\begin{equation}
  \Pi(k,\omega) = g\beta \int \frac{d^d q}{(2\pi)^d} \int \frac{d\Omega}{2\pi} G_I(k-q, \omega-\Omega) G_S(q, \Omega)  
\end{equation}

In the classical mean-field limit, the susceptible population is treated as a uniform macroscopic background, $\phi_S(x,t) = S_0$. Its momentum-space propagator is  a Dirac delta function concentrated at zero momentum and zero frequency:
\begin{equation}
    G_S^{(static)}(q, \Omega) \approx S_0 (2\pi)^{d+1} \delta^d(q) \delta(\Omega)
\end{equation}
 In the static, zero-momentum limit ($\omega \to 0, k \to 0$), this yields a simple, constant self-energy:
 \begin{equation}
     \Pi(0,0) \approx \frac{g\beta S_0}{\gamma}
 \end{equation}
 When inserted into the Dyson equation, this constant vacuum polarization generates a local mass shift for the pathogen field, defining the renormalized mass $m_R^2 = m_0^2 - \frac{g\beta S_0}{\gamma}$. 

 To describe the anomalous transmission seen in real-world data, we must restore the full dynamical kinematics of the susceptible field. If the susceptible or infected population experiences spatial diffusion (with coefficient $D_{S,I}$) alongside demographic stochasticity, their free propagators take the dynamical form:
\begin{align}
    G_S(q, \Omega)& = \frac{1}{-i\Omega + D_S q^2} \\
    G_I(k-q, \omega-\Omega)& = \frac{1}{-i(\omega-\Omega) + \gamma + D_I(k-q)^2}
\end{align}
The full vacuum polarization now requires integrating over two dynamically propagating internal lines. Substituting these into the integral yields a true convolution:
\begin{equation}
    \Pi(k,\omega) = g\beta \int \frac{d^d q}{(2\pi)^d} \int \frac{d\Omega}{2\pi} \frac{1}{-i(\omega-\Omega) + \gamma + D_I(k-q)^2} \frac{1}{-i\Omega + D_S q^2}
\end{equation}
To evaluate this integral, we first perform the contour integration over the internal frequency $\Omega$. The integrand possesses two poles in the complex $\Omega$ plane. Closing the contour in the lower half-plane for convention, the only contribution to the integral is coming from  the pole from the susceptible propagator at $\Omega = -i D_S q^2$ \cite{kamenev2011field}.Applying the Cauchy residue theorem yields the spatial momentum integral:
\begin{equation}
    \Pi(k,\omega) = g\beta \int \frac{d^d q}{(2\pi)^d} \frac{1}{-i\omega + \gamma + D_I(k-q)^2 + D_S q^2}
\end{equation}
To evaluate the $d$-dimensional momentum integral,we use the standard Feynman technics to compute one-loop diagrams \cite{peskin1995introduction}.  Expanding the denominator, we group the quadratic and linear terms of $q$:
\begin{equation}
    D_I(k^2 - 2k \cdot q + q^2) + D_S q^2 = (D_I + D_S)q^2 - 2D_I(k \cdot q) + D_I k^2
\end{equation}
 We define the total diffusion $D_T = D_I + D_S$ and the reduced diffusion $D_{red} = \frac{D_I D_S}{D_I + D_S}$, analogous to the reduced mass in a two-body physical system.
\begin{equation}
    D_T \left[ q^2 - 2 \frac{D_I}{D_T} (k \cdot q) \right] + D_I k^2 = D_T \left( q - \frac{D_I}{D_T} k \right)^2 - \frac{D_I^2}{D_T} k^2 + D_I k^2
\end{equation}
  Defining $l = q - \frac{D_I}{D_T}k$:
\begin{equation}
    \Pi(k,\omega) = g\beta \int \frac{d^d l}{(2\pi)^d} \frac{1}{D_T l^2 + \Delta(k,\omega)}
\end{equation}
where the effective mass of the loop is now strictly governed by the reduced diffusion:
\begin{equation}
    \Delta(k,\omega) = D_{red} k^2 - i\omega + \gamma
\end{equation}
using standard dimensional regularization \cite{thooft1972regularization, zinn2021quantum} to solve the spherical integral over $l$ in $d$ dimensions:
\begin{align}
    \Pi(k,\omega) &= \frac{g\beta}{D_T} \int \frac{d^d l}{(2\pi)^d} \frac{1}{l^2 + \frac{\Delta(k,\omega)}{D_T}} = \frac{g\beta}{D_T (4\pi)^{d/2}} \Gamma\left(1 - \frac{d}{2}\right) \left[ \frac{\Delta(k,\omega)}{D_T} \right]^{\frac{d}{2} - 1} \\
    & = \frac{g\beta}{(4\pi D_T)^{d/2}} \Gamma\left(1 - \frac{d}{2}\right) \left[ D_{red} k^2 - i\omega + \gamma \right]^{\frac{d}{2} - 1}
\end{align}
  the anomalous momentum scaling $(k^2)^{\frac{d}{2}-1}$ is entirely dependent on the reduced diffusion $D_{red}$. If infected individuals are perfectly quarantined ($D_I \to 0$), then $D_{red} \to 0$, and the momentum scaling drops out of the effective mass $\Delta(k,\omega)$. The fractional Riesz potential generated by the vacuum polarization \cite{metzler2000random} only emerges when both the susceptible vacuum and the infected sources are mobile \cite{colizza2006role}.

\section{Emergence of Fractional Scaling and the Dressed Propagator}

We begin with the exact $d$-dimensional dynamical vacuum polarization derived in previous section:

\begin{equation}
    \Pi(k,\omega) = \frac{g\beta}{(4\pi D_T)^{d/2}} \Gamma\left(1 - \frac{d}{2}\right) \left[ D_{red} k^2 - i\omega + \gamma \right]^{\frac{d}{2} - 1}
\end{equation}

To extract the scaling behavior that governs the active transmission phase of the epidemic, we evaluate this expression in the dynamic fluctuation regime, where the combined spatial and temporal diffusion of the hosts dominates the local, constant recovery rate ($|D_{red} k^2 - i\omega| \gg \gamma$).

We formally define the fractional scaling index $\alpha$ based on the effective dimensionality of the interaction network \cite{pastorsatorras2015epidemic}:
\begin{equation}
    \alpha = d - 2
\end{equation}
Substituting $\alpha$, the momentum and frequency-dependent contribution to the self-energy isolates into a space-time anomalous power law \cite{laskin2000fractional}:
\begin{equation}
    \Pi_{anomalous}(k, \omega) = -C_\alpha \left( D_{red} k^2 - i\omega \right)^{\alpha/2}
\end{equation}
where $$C_\alpha = - \frac{g\beta}{(4\pi D_T)^{\frac{\alpha+2}{2}}} \Gamma\left(-\frac{\alpha}{2}\right)$$

At first glance, a negative constant might appear to cause instability in the dressed pathogen propagator $D_{dressed} = (m_R^2 + C_\alpha |k|^\alpha)^{-1}$. However, the physics is perfectly preserved by the properties of the Gamma function $\Gamma(z)$ \cite{olver2010nist}. The Gamma function $\Gamma(x)$ is strictly negative for $-1 < x < 0$. Therefore, as $0 < \alpha < 2$, $\Gamma(-\alpha/2)$ inherently carries a negative sign.

\section{The Fractional Integro-Differential SIR Equations}

By integrating out the dynamically fluctuating host vacuum, the effective action of the pathogen field acquires anomalous scaling in both momentum and frequency. In this section, we apply these momentum-space field operators back to coordinate space and time to construct the modified, macroscopic integro-differential equations governing the Susceptible ($S$) and Infected ($I$) populations.

In the effective action, the pathogen field $\varphi(x,t)$ is driven by the infected population current, $I(x,t)$. The relationship is defined mathematically by the inverse of the fully dressed pathogen propagator acting upon the field.In the Fourier-Laplace domain ($k, \omega$), the equation of motion is:

\begin{equation}
    \left[ m_R^2 + C_\alpha \left( D_{red} k^2 - i\omega \right)^{\alpha/2} \right] \tilde{\varphi}(k,\omega) = g \tilde{I}(k,\omega)
\end{equation}
Consequently, the anomalous scaling strictly generates a coupled space-time fractional operator:

\begin{equation}
    m_R^2 \varphi(x,t) + C_\alpha \left( \partial_t - D_{red} \Delta \right)^{\alpha/2} \varphi(x,t) = g I(x,t)
\end{equation}
In functional analysis, the inverse fractional power of a positive differential operator $A = \partial_t - D_{red} \Delta$ is constructed using Bochner's subordination formula (the Gamma function representation of the heat semigroup \cite{bochner1949diffusion}):
\begin{equation}
    A^{-\alpha/2} I(x,t) = \frac{1}{\Gamma(\alpha/2)} \int_0^\infty s^{\frac{\alpha}{2} - 1} e^{-sA} I(x,t) \, ds
\end{equation}
The explicit definition of the inverse fractional operator acting on the infected population is the parabolic Riesz potential \cite{podlubny1998fractional}:
\begin{equation}
    \left( \partial_t - D_{red} \Delta \right)^{-\alpha/2} I(x,t) = \int_{-\infty}^t d\tau \int_{\mathbb{R}^d} d^dy \, \mathcal{K}_\alpha(x-y, t-\tau) I(y,\tau)
\end{equation}
where the fractional space-time kernel $\mathcal{K}_\alpha(x, t)$ is defined by integrating the standard heat kernel $G(x,t)$ over the auxiliary parameter $s$:
\begin{equation}
    \mathcal{K}_\alpha(x, t) = \frac{t^{\frac{\alpha}{2} - 1}}{\Gamma(\alpha/2)} G(x,t) = \frac{t^{\frac{\alpha}{2} - 1}}{\Gamma(\alpha/2)} \frac{1}{(4\pi D_{red} t)^{d/2}} \exp\left( - \frac{|x|^2}{4 D_{red} t} \right) \theta(t)
\end{equation}
where $\theta(t)$ is the Heaviside step function ensuring causality.

These results show the main properties of the QED-inspired model:

\begin{itemize}
    \item Spatial Non-Locality (The Gaussian Tail): For a fixed time difference $t-\tau$, the spatial integral smears the infection source over a distance determined by the reduced diffusion $D_{red}$.
    \item Temporal Memory (The Power-Law Decay): The term $t^{\frac{\alpha}{2} - 1}$ introduces a heavy-tailed power-law memory. Unlike standard Markovian models where previous states are "forgotten" exponentially fast via a constant decay rate $\gamma$, this fractional kernel decays algebraically.
    \item Spatiotemporal Coupling: Because the spatial Gaussian variance $4D_{red}t$ grows with the temporal delay $t$, the operator enforces a strict correlation between space and time. A pathogen shed by a super-spreader far away (large $|x-y|$) will heavily influence the susceptible population, but the peak of that influence is temporally delayed (large $t-\tau$).
\end{itemize}

In the Doi-Peliti saddle-point approximation, the classical equations for the host densities are driven by the local gauge interaction vertex, $\beta S(x,t) \varphi(x,t)$:
\begin{align}
    \frac{\partial S(x,t)}{\partial t} &= -\beta S(x,t) \varphi(x,t)\\
    \frac{\partial I(x,t)}{\partial t} & = \beta S(x,t) \varphi(x,t) - \gamma I(x,t)
\end{align}
We substitute our dynamically dressed, formal solution for the pathogen field $\varphi(x,t)$ directly into these rate equations. Defining the effective fractional transmission rate as $\beta_{eff} = \frac{\beta g}{C_\alpha}$, we obtain the complete Space-Time Fractional Integro-Differential SIR Equations:
\begin{align}
    \frac{\partial S(x,t)}{\partial t} &= -\beta_{eff} S(x,t) \left[ \left( \partial_t - D_{red} \Delta \right)^{-\alpha/2} I(x,t) \right]\\
    \frac{\partial I(x,t)}{\partial t} &= \beta_{eff} S(x,t) \left[ \left( \partial_t - D_{red} \Delta \right)^{-\alpha/2} I(x,t) \right] - \gamma I(x,t)
\end{align}
This derivation reveals a fundamental topological difference between the gauge-mediated model and existing fractional epidemic models found in the literature.

Standard approaches typically introduce fractional operators phenomenologically as linear transport terms (e.g., $-D(-\Delta)^{\alpha/2} I$) to model the anomalous movement of hosts, while leaving the $\beta S I$ transmission vertex strictly local \cite{angstmann2016fractional}.

In stark contrast, our field-theoretic derivation places the space-time fractional operator strictly inside the non-linear interaction vertex. The integration over the pathogen field proves that the transmission mechanism itself is fundamentally non-local and non-Markovian. An infection event at coordinate $x$ and time $t$ is mathematically driven by the fractional space-time integral of all infected individuals, bypassing local screening mechanisms and facilitating the explosive burst dynamics characteristic of super-spreader events.

\section{Relation to standard definition for fractional derivatives }
Standard fractional epidemic models usually inject separate, decoupled operators: the Riemann-Liouville (or Caputo) derivative for time, and the Riesz fractional Laplacian for space \cite{metzler2000random, lu2023anomalous, podlubny1998fractional}. 

\subsection{The Temporal Limit: Recovering the Riemann-Liouville Derivative}

To keep only the temporal memory effects, we consider a regime where spatial diffusion is negligible or the system is spatially homogeneous. In this limit, the spatial Laplacian vanishes ($\Delta \to 0$) and one gets:
\begin{equation}
    \left( \partial_t - D_{red} \Delta \right)^{-\alpha/2} \xrightarrow{D_{red} \to 0} (\partial_t)^{-\alpha/2}
\end{equation}
As $D_{red} \to 0$, the spatial heat kernel $G(x-y, t-\tau)$ collapses strictly into a Dirac delta function, $\delta^d(x-y)$, enforcing absolute spatial locality \cite{evans2010partial}. The spatial integral disappears, leaving only the temporal convolution:
\begin{equation}
    (\partial_t)^{-\alpha/2} I(x,t) = \frac{1}{\Gamma(\alpha/2)} \int_0^t (t-\tau)^{\frac{\alpha}{2} - 1} I(x,\tau) \, d\tau
\end{equation}
where we can recognize the Riemann-Liouville fractional integral of order $\alpha/2$ \cite{kilbas2006theory}. Inversing this operator, one obtains the standard Riemann-Liouville or Caputo fractional time derivatives ($\partial_t^{\alpha/2}$) used universally in literature to model non-Markovian memory and incubation delays \cite{angstmann2016fractional, saeedian2017memory}.

\subsection{The Spatial Limit: Recovering the Riesz Fractional Laplacian}

Conversely, to isolate the spatial non-locality (Lévy flights), we look at the stationary state or the long-time asymptotic limit of the epidemic where temporal fluctuations average out ($\partial_t \to 0$).The operator reduces to a pure spatial operator:

\begin{equation}
    \left( \partial_t - D_{red} \Delta \right)^{-\alpha/2} \xrightarrow{\partial_t \to 0} (-D_{red} \Delta)^{-\alpha/2}
\end{equation}
Computing it, one gets:
\begin{equation}
    (-D_{red} \Delta)^{-\alpha/2} I(x) \propto \int_{\mathbb{R}^d} \frac{I(y)}{|x-y|^{d-\alpha}} \, d^dy
\end{equation}
This is the exact, standard definition of the Riesz potential, which is the mathematical inverse of the Fractional Laplacian $(-\Delta)^{\alpha/2}$ \cite{lischke2020what, pozrikidis2016fractional}. This operator is used to model spatial super-spreaders and infinite-variance random walks (Lévy flights) in  literature \cite{metzler2000random}.

\section{Computation of the  Effective Reproductive Number ($R_{eff}(k,\omega)$)}

In the standard Doi-Peliti mean-field approximation, the base reproductive number is obtained  from the zero-mode (static, infinite wavelength) limit of the transmission vertex \cite{diekmann1990definition}:
\begin{equation}
    R_0 = \frac{g\beta S_0}{\gamma m_0^2}
\end{equation}
However, the true transmission rate between a susceptible host at space-time coordinate $(x,t)$ and an infected host at $(y,\tau)$ is mediated by the fully dressed pathogen propagator \cite{tauber2014critical}. In the Fourier-Laplace domain, the effective macroscopic interaction potential is strictly proportional to this dressed propagator:

\begin{align}
    V_{eff}(k, \omega) &= g\beta D_{dressed}(k, \omega)\\
 & = \frac{g\beta}{m_R^2 + C_\alpha (D_{red} k^2 - i\omega)^{\alpha/2}}
\end{align}
%To find the spectral reproductive number $R_{eff}(k, \omega)$, we scale this effective interaction by the available susceptible fraction $S(t)$ and the recovery rate $\gamma$, just as in the classical derivation:
Using the standard definition for the effective reproductive number $R_{eff}$, one gets:
\begin{align}
    R_{eff}(k, \omega) &= \frac{S(t)}{\gamma} V_{eff}(k, \omega) = \frac{g\beta S(t)}{\gamma} \left[ \frac{1}{m_R^2 + C_\alpha (D_{red} k^2 - i\omega)^{\alpha/2}} \right]\\
    & = R_0 \frac{S(t)}{S_0} \left[ \frac{m_0^2}{m_R^2 + C_\alpha (D_{red} k^2 - i\omega)^{\alpha/2}} \right]
\end{align}
This solution can be understood according the following interpretation:
\begin{itemize}
    \item The Real Part ($\text{Re}[R_{eff}]$) - Direct Contagion: This represents the in-phase transmission. It dictates the secondary infections that occur immediately and directly as the pathogen diffuses spatially. It governs the structural size of the outbreak clusters (dependent on $k$).
    \item The Imaginary Part ($\text{Im}[R_{eff}]$) - Retarded Contagion (Memory):This represents the out-of-phase transmission. It is strictly driven by $\omega$ \cite{chaikin1995principles}. An imaginary reproduction number proves that secondary infections are phase-shifted (delayed) from the primary cases. This mathematical phase shift perfectly encodes the biological incubation period and the lingering environmental persistence of the pathogen.
\end{itemize}
To understand better this form to define the $R_{eff}$, it is interesting to take the asymptotic limits:
\begin{itemize}
    \item The Endemic/Macro Limit ($k \to 0, \omega \to 0$): If we look at the whole country over a long period, the dynamic terms vanish. $R_{eff}$ returns to the real, classical scalar $\approx R_0 \frac{S}{S_0} (\frac{m_0^2}{m_R^2})$.
    \item The High-Frequency Burst Limit ($\omega \gg D_{red} k^2$):During a rapid super-spreader event (a frat party, a crowded concert), the frequency of transmission is extremely high. The $R_{eff}$ is dominated by $(-i\omega)^{\alpha/2}$. The transmission is heavily phase-shifted, meaning the massive wave of secondary infections will not become visible until after a specific time delay.
    \item The High-Wavelength Cluster Limit ($k \gg \omega$): If the disease is spreading slowly but forming tight spatial clusters, $R_{eff}$ is dominated by $C_\alpha (D_{red} k^2)^{\alpha/2}$. The transmission capability strictly depends on the reduced diffusion of the hosts.
\end{itemize}
A pathogen might have an $R_{eff} < 1$ for slow, macro-scale endemic spread (meaning the pandemic is "technically" over at the country level), while simultaneously having an $R_{eff} \gg 1$ for high-frequency, localized bursts.
Let us to study the both limit cases.To establish the complete theoretical behavior of the epidemic near the phase transition, we must evaluate the macroscopic Effective Reproductive Number, $R_{eff}$, in its two extreme asymptotic limits.

By starting from the fully dressed spectral propagator, we can compute the 1-loop vertex renormalization for both the static limit (dominated by spatial clusters) and the high-frequency limit (dominated by temporal bursts).

The macroscopic transmission rate is governed by the 1-loop vertex renormalization, which integrates the fully dressed pathogen propagator over all internal momenta $k$ up to the ultraviolet cutoff $\Lambda$ \cite{binney1992theory}:
\begin{equation}
    \beta_{ren}(\omega) = \beta_0 \left( 1 - \frac{g\beta_0}{\gamma} \int \frac{d^d k}{(2\pi)^d} \frac{1}{m_R^2 + C_\alpha (D_{red} k^2 - i\omega)^{\alpha/2}} \right)
\end{equation}
We evaluate this integral at the critical threshold of the epidemic ($m_R \to 0$) to see exactly how the anomalous scaling governs the phase transition.

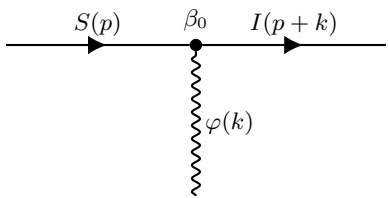
\begin{figure}[htbp]
    \centering
    \begin{tikzpicture}
        % Vertices
        \coordinate (S_in) at (0, 0);
        \node[circle, draw=black, fill=black, inner sep=1.5pt, label=above:{$\beta_0$}] (V1) at (2.5, 0) {};
        \coordinate (I_out) at (5, 0);
        \coordinate (Phi_in) at (2.5, -2);

        % Lines
        \draw[draw=black, thick, postaction={decorate}, decoration={markings, mark=at position 0.55 with {\arrow{Latex[length=2.5mm, width=2.5mm]}}}] (S_in) -- node[above] {$S(p)$} (V1);
        \draw[draw=black, thick, postaction={decorate}, decoration={markings, mark=at position 0.55 with {\arrow{Latex[length=2.5mm, width=2.5mm]}}}] (V1) -- node[above] {$I(p+k)$} (I_out);
        \draw[draw=black, thick, decorate, decoration={snake, amplitude=1.5pt, segment length=5pt}] (Phi_in) -- node[right] {$\varphi(k)$} (V1);
    \end{tikzpicture}
    \caption{Tree-level Feynman diagram representing the fundamental transmission vertex. A susceptible host field $S$ with momentum $p$ absorbs a quantum of the pathogen gauge field $\varphi$ carrying momentum $k$, transitioning into the infected state $I$ with conserved momentum $p+k$. The bare coupling constant for this local interaction is the transmission rate $\beta_0$.}
    \label{fig:vertex_beta}
\end{figure}

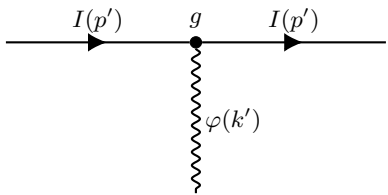
\begin{figure}[htbp]
    \centering
    \begin{tikzpicture}
        % Vertices
        \coordinate (I_in) at (0, 0);
        \node[circle, draw=black, fill=black, inner sep=1.5pt, label=above:{$g$}] (V2) at (2.5, 0) {};
        \coordinate (I_out) at (5, 0);
        \coordinate (Phi_out) at (2.5, -2);

        % Lines
        \draw[draw=black, thick, postaction={decorate}, decoration={markings, mark=at position 0.55 with {\arrow{Latex[length=2.5mm, width=2.5mm]}}}] (I_in) -- node[above] {$I(p')$} (V2);
        \draw[draw=black, thick, postaction={decorate}, decoration={markings, mark=at position 0.55 with {\arrow{Latex[length=2.5mm, width=2.5mm]}}}] (V2) -- node[above] {$I(p')$} (I_out);
        \draw[draw=black, thick, decorate, decoration={snake, amplitude=1.5pt, segment length=5pt}] (V2) -- node[right] {$\varphi(k')$} (Phi_out);
    \end{tikzpicture}
    \caption{Tree-level Feynman diagram of the pathogen emission vertex. The infected host population $I$ acts as the dynamical source current for the mediating gauge field, shedding the pathogen $\varphi$ into the environment. The strength of this shedding mechanism is governed by the bare emission coupling $g$.}
    \label{fig:vertex_g}
\end{figure}

% ==========================================
% DIAGRAM 3: 1-Loop Vertex Correction (\beta_{ren})
% ==========================================
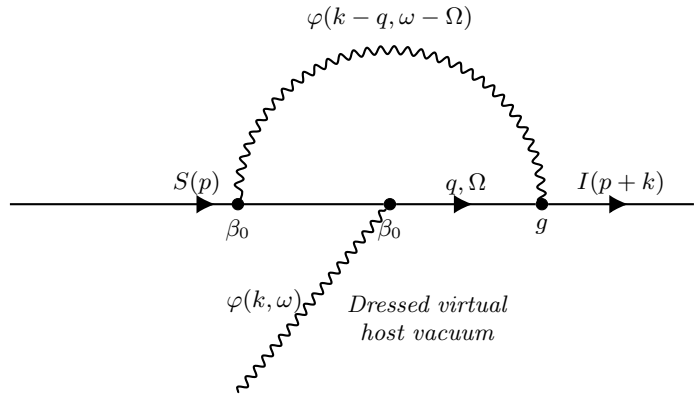
\begin{figure}[htbp]
    \centering
\begin{tikzpicture}
    % External coordinates
    \coordinate (S_in) at (-1, 0);
    \coordinate (I_out) at (8, 0);
    \coordinate (Phi_ext) at (2, -2.5);

    % Interaction vertices
    \node[vertex, label=below:{$\beta_0$}] (V_beta) at (4, 0) {};
    \node[vertex, label=below:{$g$}] (V_g) at (6, 0) {};
     \node[vertex, label=below:{$\beta_0$}]  at (2, 0) {};

    % Host lines (straight)
    \draw[fermion] (S_in) -- node[ above] {$S(p)$} (V_beta);
    \draw[fermion] (V_beta) -- node[above] {$q, \Omega$} (V_g); % Virtual host loop
    \draw[fermion] (V_g) -- node[above] {$I(p+k)$} (I_out);

    % External pathogen field
    \draw[gauge] (Phi_ext) -- node[left] {$\varphi(k, \omega)$} (V_beta);

    % Virtual pathogen loop (semi-circle)
    % Draws an arc from (6,0) to (2,0) with a radius of 2
    \draw[gauge] (V_g) arc[start angle=0, end angle=180, radius=2cm];
    
    % Label for the virtual pathogen loop
    \node[above] at (4, 2.2) {$\varphi(k-q, \omega-\Omega)$};
    
    % Optional descriptive text
    \node[align=center, text width=4cm] at (4.5, -1.5) {\small \textit{Dressed virtual}\\ \textit{host vacuum}};
\end{tikzpicture}
  \caption{Tree-level Feynman diagram of the pathogen emission vertex. The infected host population $I$ acts as the dynamical source current for the mediating gauge field, shedding the pathogen $\varphi$ into the environment. The strength of this shedding mechanism is governed by the bare emission coupling $g$.}
    \label{fig:vertex_g}
    \end{figure}

\subsection{ Static limit ($\omega \to 0$): Condition: $D_{red} k^2 \gg |\omega|$}

This regime applies to the long-range, quasi-stationary spread of the disease. The temporal fluctuations are slow compared to the spatial diffusion of the pathogen.

Dropping the $\omega$ term, the denominator is governed purely by the spatial fractional scaling $k^\alpha$. As we derived previously, we substitute the topological relation $\alpha = d - 2$ and evaluate the $d$-dimensional integral over the spherical coordinates:
\begin{equation}
    I_{static} = \frac{S_d}{(2\pi)^d C_\alpha D_{red}^{\alpha/2}} \int_0^\Lambda \frac{k^{d-1}}{k^\alpha} dk = \frac{S_d}{(2\pi)^d \tilde{C}_\alpha} \int_0^\Lambda k^1 dk = \frac{S_d}{(2\pi)^d \tilde{C}_\alpha} \frac{\Lambda^2}{2}
\end{equation}

Substituting this back into the vertex correction and multiplying by the susceptible fraction yields the static fractal $R_{eff}$:
\begin{equation}
    R_{eff}^{static} = R_0 \frac{S(t)}{S_0} \left[ 1 - \lambda_{static} \Lambda^2 \right]
\end{equation}
where $\lambda_{static} = \frac{g \beta_0 S_d}{2 \gamma (2\pi)^d \tilde{C}_\alpha}$
The macroscopic transmission becomes strictly bound by $\Lambda^2$. Because $\Lambda$ represents the inverse of the minimum interaction distance ($1/a$), the mathematics prove that a Lévy-flight-driven epidemic can only be suppressed by altering the microscopic spatial geometry of the network (e.g., strict minimum-distance social distancing) \cite{hufnagel2004forecast}. Because $m_R \to 0$, the correlation length of the outbreak diverges \cite{goldenfeld1992lectures}. The system loses its characteristic cluster size, and the remaining dressed propagator isolates to a pure, scale-invariant power law: $D_{dressed}(k) \propto 1/k^\alpha$.

From these results, we  can compute  the spatial interaction probability $P(r)$ as a function of the distance $r$ between hosts \cite{metzler2000random}:
\begin{equation}P(r) \propto \frac{1}{r^{d-\alpha}}
\end{equation}
This is the  mathematical signature of a Lévy flight.  The Lévy flight distribution ($1/r^{d-\alpha}$) generated by the QED-SIR field theory proves there is a persistent, non-negligible probability of an infected host making a massive spatial jump (e.g., boarding an airplane or attending a mass gathering) and seeding a new cluster far beyond the local transmission zone \cite{brockmann2006scaling}. That new cluster grows locally, but eventually, the heavy-tailed variance guarantees another massive spatial jump will occur. Geometrically, this non-local transmission network constructs a fractal topology—clusters within clusters separated by vast distances—occurring at all length scales simultaneously \cite{song2005self}.
\subsection{ The High-Frequency Limit ($|\omega| \gg D_{red} k^2$)}

This regime correspond to rapid burst events (e.g., a super-spreader event in a confined space) \cite{althouse2020superspreading}. Spatial diffusion is negligible compared to the extreme rate of temporal interaction, so we get:
\begin{equation}
    I_{dynamic}(\omega) = \frac{S_d}{(2\pi)^d C_\alpha (-i\omega)^{\alpha/2}} \int_0^\Lambda k^{d-1} dk = \frac{S_d}{(2\pi)^d C_\alpha (-i\omega)^{\alpha/2}} \frac{\Lambda^d}{d}
\end{equation}
The  frequency-dependent transmission rate is now given by:
\begin{equation}
    \beta_{ren}(\omega) = \beta_0 \left[ 1 - \lambda_{dynamic} \Lambda^d (-i\omega)^{-\alpha/2} \right]
\end{equation}
where $\lambda_{dynamic} = \frac{g \beta_0 S_d}{d \gamma (2\pi)^d C_\alpha}$. The multiplier $(-i\omega)^{-\alpha/2}$ transforms into the Riemann-Liouville fractional integral operator, $\partial_t^{-\alpha/2}$.
Therefore, in the time domain, the effective reproductive number is given by:
\begin{equation}
    R_{eff}^{dynamic}(t) = R_0 \frac{S(t)}{S_0} \left[ 1 - \lambda_{dynamic} \Lambda^d \partial_t^{-\alpha/2} \right]
\end{equation}
where $\Lambda^d$ is the interaction volume. The fractional operator $\partial_t^{-\alpha/2}$ acts as a convolution over the entire history of the outbreak, weighted by a heavy-tailed power-law decay $(t-\tau)^{\frac{\alpha}{2}-1}$ \cite{podlubny1998fractional}.
To see the effect of $S$ and $I$ equations, one has to use that:
\begin{equation}
    \tilde{\varphi}(\omega) = \frac{g}{C_\alpha} (-i\omega)^{-\alpha/2} \tilde{I}(\omega)
\end{equation}
 In fractional calculus, the multiplier $(-i\omega)^{-\alpha/2}$ is the exact spectral definition of the Riemann-Liouville fractional integral,  written as $\partial_t^{-\alpha/2}$. 
In coordinate time, the pathogen field becomes a convolution with a heavy-tailed power law:
\begin{equation}
\varphi(t) = \frac{g}{C_\alpha} \frac{1}{\Gamma(\alpha/2)} \int_0^t (t-\tau)^{\frac{\alpha}{2}-1} I(\tau) \, d\tau
\end{equation}
Defining the effective coupling $\beta_{eff} = \frac{\beta g}{C_\alpha}$, we obtain the modified, high-wavelength macroscopic equations:
\begin{align}
    \frac{dS(t)}{dt} &= -\frac{\beta_{eff}}{\Gamma(\alpha/2)} S(t) \int_0^t (t-\tau)^{\frac{\alpha}{2}-1} I(\tau) \, d\tau \\
    \frac{dI(t)}{dt}& = \frac{\beta_{eff}}{\Gamma(\alpha/2)} S(t) \int_0^t (t-\tau)^{\frac{\alpha}{2}-1} I(\tau) \, d\tau - \gamma I(t)
\end{align}
%To fully grasp the epidemiological consequences of the dynamic, high-wavelength limit ($k \to 0$), we must contrast the fractional integro-differential rate equations with the standard SIR differential equations.
In the classical, adiabatic approximation of the Doi-Peliti framework, the pathogen field is assumed to equilibrate instantaneously with the host population ($\partial_t \varphi \approx 0$). Mathematically, this forces the temporal interaction kernel to collapse into a simple Dirac delta function, $K(t-\tau) = \delta(t-\tau)$. Consequently, the instantaneous infection rate depends on $I(t)$. This forces a Markovian (memoryless) physical reality: it assumes that biological residence times in the infectious compartment follow a perfectly exponential distribution, incubation periods are negligible, and the pathogen vanishes from the environment the moment a host recovers.However, by  evaluating the dynamical loop integral in the anomalous regime ($0 < \alpha < 2$), our derivation proves that integrating out a fluctuating, heterogeneous host vacuum mathematically forbids this instantaneous equilibration. The resulting  operator generates a temporal convolution with a  heavy-tailed power-law kernel: $K(t-\tau) \propto (t-\tau)^{\frac{\alpha}{2}-1}$. Now, the macroscopic mechanics of transmission are given through three distinct physical effects:
\begin{itemize}
    \item Non-Exponential Incubation and Shedding: Because $I(t)$  is now a weighted integral over the  curve $I(\tau), \tau \ll t$, the model  accounts for heavy-tailed waiting-time distributions %This captures the reality of highly variable, non-exponential incubation periods, where secondary infections are triggered by primary cases that were exposed days or weeks prior 
    \cite{lloyd2001realistic}.
    \item Environmental Reservoirs: The slow, algebraic decay of the $(t-\tau)^{\frac{\alpha}{2}-1}$ kernel shows that  the pathogen field  continue to drive secondary infections long after the initial shedding event has ceased, even if the instantaneous count $I(t)$ appears negligible \cite{morawska2020time}.
    \item Temporal Avalanches: Because classical SIR equations are governed by local, instantaneous derivatives, they  predict smooth, continuous curves. Conversely, our fractional model dictates that transmission is mathematically phase-shifted. The system silently accumulates historical transmission potential via the fractional integral. This is generating periods of latency followed by sudden, explosive temporal avalanches—bursts of new cases \cite{barabasi2005origin}.
\end{itemize}
\section{Epidemiological Consequences}

By going beyond the mean-field approximation, our fractional integro-differential model overturns several classical epidemiological assumptions.

\subsection{The Fallacy of Local Herd Immunity}
Classical SIR models predict that the accumulation of individuals which recovered from the disease and are immunized creates a local shield.  However, we have shown that the Lévy flight distribution ($1/r^{d-\alpha}$) ensures that an infected host possesses a persistent probability of making a massive spatial jump. Because the pathogen can leap over local zones of immunity to seed new clusters far beyond the local transmission zone, the local herd immunity fails to contain the macroscopic spread \cite{riley2007large}. Geometrically, this results in a non-local transmission network forming a fractal topology of clusters within clusters.

\subsection{Microscopic Constraints on Macroscopic Spread}
Because the local screening mechanisms fail, the static effective reproductive number ($R_{eff}^{static}$) becomes  bounded by the ultraviolet cutoff, $\Lambda^2$ which represents the inverse of the minimum microscopic interaction distance between hosts ($1/a$)\cite{glass2006targeted}. It shows that  a super-spreader-driven epidemic can only be collapsed by  altering the microscopic spatial geometry of the network, such as enforcing strict minimum-distance social distancing protocols \cite{glass2006targeted}. 

\subsection{Warning Signs of Avalanches}
Evaluating the high-frequency limit ($|\omega| \gg D_{red} k^2$) reveals the danger of interpreting short-term case drops as the end of an outbreak. The fractional time operator is storing ``epidemic charge''. The pathogen lingers in the environment or persists in asymptomatic super-spreaders, actively driving secondary infections long after the initial shedding event \cite{morawska2020time}. Consequently, the system silently accumulates historical transmission potential. It will exhibit phase-shifted burst dynamics, characterized by periods of apparent dormancy followed by sudden, explosive temporal avalanches \cite{barabasi2005origin}.

\subsection{Mobility effects}
The fractional Riesz potential is  governed by the reduced diffusion of the host network, $D_{red}$. The anomalous momentum scaling $(k^2)^{\frac{d}{2}-1}$ is  dependent on this term. If infected individuals are perfectly quarantined ($D_I \to 0$), the reduced diffusion vanishes ($D_{red} \to 0$), and the anomalous momentum scaling drops  out of the effective mass \cite{eubank2004modelling}. This proves that fractional, non-local transmission leaps only emerge when both the susceptible vacuum and the infected sources are mobile. Thus, rapid, targeted isolation of highly infectious individuals is mathematically required to collapse the fractional scaling back into manageable, localized Brownian diffusion.

\section{Discussion and Conclusion}
The  limitation of classical compartmental models lies in their assumption of instantaneous, memoryless equilibration. By applying the Doi-Peliti formalism to model epidemic dynamics as a gauge-mediated field theory, we have established a  mathematical connection between stochastic micro-foundations to macroscopic fractional spectral decomposition.

Our derivation proves that integrating out a fluctuating, heterogeneous host vacuum mathematically forbids instantaneous equilibration. Instead, the transmission mechanism itself is fundamentally non-local and non-Markovian. We have shown that the coupled parabolic operator we derived is actually the mathematical parent of both the Riemann-Liouville derivative and the Riesz fractional Laplacian.

This spectral approach  redefines the phase transition of a pandemic. In the static limit, the mathematics prove that a Lévy-flight-driven epidemic can only be suppressed by altering the microscopic spatial geometry of the network (e.g., strict minimum-distance social distancing). Furthermore, the temporal dynamics reveal an ``epidemic capacitor'' effect: the slow, algebraic decay of the interaction kernel dictates that the pathogen lingers in the environment or persists in asymptomatic super-spreaders, actively continuing to drive secondary infections long after the initial shedding event has ceased, even if the instantaneous count appears negligible. The system silently accumulates historical transmission potential via the fractional integral , creating periods of apparent dormancy followed by sudden, explosive temporal avalanches.

\begin{acknowledgments}
We acknowledge financial support from SECIHTI and SNII (M\'exico) and Guanajuato University.
\end{acknowledgments}

%\bibliographystyle{apsrev4-2}
%\bibliography{ref.bib}

\begin{thebibliography}{41}%
\makeatletter
\providecommand \@ifxundefined [1]{%
 \@ifx{#1\undefined}
}%
\providecommand \@ifnum [1]{%
 \ifnum #1\expandafter \@firstoftwo
 \else \expandafter \@secondoftwo
 \fi
}%
\providecommand \@ifx [1]{%
 \ifx #1\expandafter \@firstoftwo
 \else \expandafter \@secondoftwo
 \fi
}%
\providecommand \natexlab [1]{#1}%
\providecommand \enquote  [1]{``#1''}%
\providecommand \bibnamefont  [1]{#1}%
\providecommand \bibfnamefont [1]{#1}%
\providecommand \citenamefont [1]{#1}%
\providecommand \href@noop [0]{\@secondoftwo}%
\providecommand \href [0]{\begingroup \@sanitize@url \@href}%
\providecommand \@href[1]{\@@startlink{#1}\@@href}%
\providecommand \@@href[1]{\endgroup#1\@@endlink}%
\providecommand \@sanitize@url [0]{\catcode `\\12\catcode `\$12\catcode `\&12\catcode `\#12\catcode `\^12\catcode `\_12\catcode `\%12\relax}%
\providecommand \@@startlink[1]{}%
\providecommand \@@endlink[0]{}%
\providecommand \url  [0]{\begingroup\@sanitize@url \@url }%
\providecommand \@url [1]{\endgroup\@href {#1}{\urlprefix }}%
\providecommand \urlprefix  [0]{URL }%
\providecommand \Eprint [0]{\href }%
\providecommand \doibase [0]{https://doi.org/}%
\providecommand \selectlanguage [0]{\@gobble}%
\providecommand \bibinfo  [0]{\@secondoftwo}%
\providecommand \bibfield  [0]{\@secondoftwo}%
\providecommand \translation [1]{[#1]}%
\providecommand \BibitemOpen [0]{}%
\providecommand \bibitemStop [0]{}%
\providecommand \bibitemNoStop [0]{.\EOS\space}%
\providecommand \EOS [0]{\spacefactor3000\relax}%
\providecommand \BibitemShut  [1]{\csname bibitem#1\endcsname}%
\let\auto@bib@innerbib\@empty
%</preamble>
\bibitem [{\citenamefont {Kermack}\ and\ \citenamefont {McKendrick}(1927)}]{kermack1927contribution}%
  \BibitemOpen
  \bibfield  {author} {\bibinfo {author} {\bibfnamefont {W.~O.}\ \bibnamefont {Kermack}}\ and\ \bibinfo {author} {\bibfnamefont {A.~G.}\ \bibnamefont {McKendrick}},\ }\href {https://doi.org/10.1098/rspa.1927.0118} {\bibfield  {journal} {\bibinfo  {journal} {Proceedings of the Royal Society of London. Series A, Containing papers of a mathematical and physical character}\ }\textbf {\bibinfo {volume} {115}},\ \bibinfo {pages} {700} (\bibinfo {year} {1927})}\BibitemShut {NoStop}%
\bibitem [{\citenamefont {Lloyd-Smith}\ \emph {et~al.}(2005)\citenamefont {Lloyd-Smith}, \citenamefont {Schreiber}, \citenamefont {Kopp},\ and\ \citenamefont {Getz}}]{lloyd2005superspreading}%
  \BibitemOpen
  \bibfield  {author} {\bibinfo {author} {\bibfnamefont {J.~O.}\ \bibnamefont {Lloyd-Smith}}, \bibinfo {author} {\bibfnamefont {S.~J.}\ \bibnamefont {Schreiber}}, \bibinfo {author} {\bibfnamefont {P.~E.}\ \bibnamefont {Kopp}},\ and\ \bibinfo {author} {\bibfnamefont {W.~M.}\ \bibnamefont {Getz}},\ }\href {https://doi.org/10.1038/nature04153} {\bibfield  {journal} {\bibinfo  {journal} {Nature}\ }\textbf {\bibinfo {volume} {438}},\ \bibinfo {pages} {355} (\bibinfo {year} {2005})}\BibitemShut {NoStop}%
\bibitem [{\citenamefont {Brockmann}\ \emph {et~al.}(2006)\citenamefont {Brockmann}, \citenamefont {Hufnagel},\ and\ \citenamefont {Geisel}}]{brockmann2006scaling}%
  \BibitemOpen
  \bibfield  {author} {\bibinfo {author} {\bibfnamefont {D.}~\bibnamefont {Brockmann}}, \bibinfo {author} {\bibfnamefont {L.}~\bibnamefont {Hufnagel}},\ and\ \bibinfo {author} {\bibfnamefont {T.}~\bibnamefont {Geisel}},\ }\href {https://doi.org/10.1038/nature04292} {\bibfield  {journal} {\bibinfo  {journal} {Nature}\ }\textbf {\bibinfo {volume} {439}},\ \bibinfo {pages} {462} (\bibinfo {year} {2006})}\BibitemShut {NoStop}%
\bibitem [{\citenamefont {Lu}\ \emph {et~al.}(2023)\citenamefont {Lu}, \citenamefont {Ren}, \citenamefont {Chen}, \citenamefont {Meng},\ and\ \citenamefont {Yu}}]{lu2023anomalous}%
  \BibitemOpen
  \bibfield  {author} {\bibinfo {author} {\bibfnamefont {Z.}~\bibnamefont {Lu}}, \bibinfo {author} {\bibfnamefont {G.}~\bibnamefont {Ren}}, \bibinfo {author} {\bibfnamefont {Y.}~\bibnamefont {Chen}}, \bibinfo {author} {\bibfnamefont {X.}~\bibnamefont {Meng}},\ and\ \bibinfo {author} {\bibfnamefont {Y.}~\bibnamefont {Yu}},\ }\href {https://doi.org/10.1142/S1793524522501303} {\bibfield  {journal} {\bibinfo  {journal} {International Journal of Biomathematics}\ }\textbf {\bibinfo {volume} {16}},\ \bibinfo {pages} {2250130} (\bibinfo {year} {2023})}\BibitemShut {NoStop}%
\bibitem [{\citenamefont {Yang}\ \emph {et~al.}(2025)\citenamefont {Yang}, \citenamefont {Yu},\ and\ \citenamefont {Shi}}]{yang2025fractional}%
  \BibitemOpen
  \bibfield  {author} {\bibinfo {author} {\bibfnamefont {J.-W.}\ \bibnamefont {Yang}}, \bibinfo {author} {\bibfnamefont {Z.-G.}\ \bibnamefont {Yu}},\ and\ \bibinfo {author} {\bibfnamefont {L.}~\bibnamefont {Shi}},\ }\href {https://doi.org/10.1063/5.0288024} {\bibfield  {journal} {\bibinfo  {journal} {Chaos: An Interdisciplinary Journal of Nonlinear Science}\ }\textbf {\bibinfo {volume} {35}},\ \bibinfo {pages} {113112} (\bibinfo {year} {2025})}\BibitemShut {NoStop}%
\bibitem [{\citenamefont {Doi}(1976)}]{doi1976second}%
  \BibitemOpen
  \bibfield  {author} {\bibinfo {author} {\bibfnamefont {M.}~\bibnamefont {Doi}},\ }\href {https://doi.org/10.1088/0305-4470/9/9/008} {\bibfield  {journal} {\bibinfo  {journal} {Journal of Physics A: Mathematical and General}\ }\textbf {\bibinfo {volume} {9}},\ \bibinfo {pages} {1465} (\bibinfo {year} {1976})}\BibitemShut {NoStop}%
\bibitem [{\citenamefont {Peliti}(1985)}]{peliti1985path}%
  \BibitemOpen
  \bibfield  {author} {\bibinfo {author} {\bibfnamefont {L.}~\bibnamefont {Peliti}},\ }\href {https://doi.org/10.1051/jphys:019850046090146900} {\bibfield  {journal} {\bibinfo  {journal} {Journal de Physique}\ }\textbf {\bibinfo {volume} {46}},\ \bibinfo {pages} {1469} (\bibinfo {year} {1985})}\BibitemShut {NoStop}%
\bibitem [{\citenamefont {T{\"a}uber}(2014)}]{tauber2014critical}%
  \BibitemOpen
  \bibfield  {author} {\bibinfo {author} {\bibfnamefont {U.~C.}\ \bibnamefont {T{\"a}uber}},\ }\href {https://doi.org/10.1017/CBO9781139046213} {\emph {\bibinfo {title} {Critical Dynamics: A Field Theory Approach to Equilibrium and Non-Equilibrium Scaling Behavior}}}\ (\bibinfo  {publisher} {Cambridge University Press},\ \bibinfo {year} {2014})\BibitemShut {NoStop}%
\bibitem [{\citenamefont {Bernal-Alvarado}\ and\ \citenamefont {Delepine}(2026)}]{bernalalvarado2026gaugemediated}%
  \BibitemOpen
  \bibfield  {author} {\bibinfo {author} {\bibfnamefont {J.~d.~J.}\ \bibnamefont {Bernal-Alvarado}}\ and\ \bibinfo {author} {\bibfnamefont {D.}~\bibnamefont {Delepine}},\ }\bibfield  {journal} {\bibinfo  {journal} {arXiv preprint arXiv:2602.13913}\ }\href {https://doi.org/10.48550/arXiv.2602.13913} {10.48550/arXiv.2602.13913} (\bibinfo {year} {2026})\BibitemShut {NoStop}%
\bibitem [{\citenamefont {Cardy}\ and\ \citenamefont {Grassberger}(1985)}]{cardy1985epidemic}%
  \BibitemOpen
  \bibfield  {author} {\bibinfo {author} {\bibfnamefont {J.~L.}\ \bibnamefont {Cardy}}\ and\ \bibinfo {author} {\bibfnamefont {P.}~\bibnamefont {Grassberger}},\ }\href {https://doi.org/10.1088/0305-4470/18/6/001} {\bibfield  {journal} {\bibinfo  {journal} {Journal of Physics A: Mathematical and General}\ }\textbf {\bibinfo {volume} {18}},\ \bibinfo {pages} {L267} (\bibinfo {year} {1985})}\BibitemShut {NoStop}%
\bibitem [{\citenamefont {Metzler}\ and\ \citenamefont {Klafter}(2000)}]{metzler2000random}%
  \BibitemOpen
  \bibfield  {author} {\bibinfo {author} {\bibfnamefont {R.}~\bibnamefont {Metzler}}\ and\ \bibinfo {author} {\bibfnamefont {J.}~\bibnamefont {Klafter}},\ }\href {https://doi.org/10.1016/S0370-1573(00)00070-3} {\bibfield  {journal} {\bibinfo  {journal} {Physics Reports}\ }\textbf {\bibinfo {volume} {339}},\ \bibinfo {pages} {1} (\bibinfo {year} {2000})}\BibitemShut {NoStop}%
\bibitem [{\citenamefont {Janssen}\ and\ \citenamefont {T{\"a}uber}(2005)}]{janssen2005field}%
  \BibitemOpen
  \bibfield  {author} {\bibinfo {author} {\bibfnamefont {H.-K.}\ \bibnamefont {Janssen}}\ and\ \bibinfo {author} {\bibfnamefont {U.~C.}\ \bibnamefont {T{\"a}uber}},\ }\href {https://doi.org/10.1016/j.aop.2004.09.011} {\bibfield  {journal} {\bibinfo  {journal} {Annals of Physics}\ }\textbf {\bibinfo {volume} {315}},\ \bibinfo {pages} {147} (\bibinfo {year} {2005})}\BibitemShut {NoStop}%
\bibitem [{\citenamefont {Kamenev}(2011)}]{kamenev2011field}%
  \BibitemOpen
  \bibfield  {author} {\bibinfo {author} {\bibfnamefont {A.}~\bibnamefont {Kamenev}},\ }\href {https://doi.org/10.1017/CBO9781139003667} {\emph {\bibinfo {title} {Field Theory of Non-Equilibrium Systems}}}\ (\bibinfo  {publisher} {Cambridge University Press},\ \bibinfo {year} {2011})\BibitemShut {NoStop}%
\bibitem [{\citenamefont {Peskin}\ and\ \citenamefont {Schroeder}(1995)}]{peskin1995introduction}%
  \BibitemOpen
  \bibfield  {author} {\bibinfo {author} {\bibfnamefont {M.~E.}\ \bibnamefont {Peskin}}\ and\ \bibinfo {author} {\bibfnamefont {D.~V.}\ \bibnamefont {Schroeder}},\ }\href {https://doi.org/10.1201/9780429503559} {\emph {\bibinfo {title} {An Introduction to Quantum Field Theory}}}\ (\bibinfo  {publisher} {CRC Press},\ \bibinfo {year} {1995})\BibitemShut {NoStop}%
\bibitem [{\citenamefont {'t~Hooft}\ and\ \citenamefont {Veltman}(1972)}]{thooft1972regularization}%
  \BibitemOpen
  \bibfield  {author} {\bibinfo {author} {\bibfnamefont {G.}~\bibnamefont {'t~Hooft}}\ and\ \bibinfo {author} {\bibfnamefont {M.}~\bibnamefont {Veltman}},\ }\href {https://doi.org/10.1016/0550-3213(72)90279-9} {\bibfield  {journal} {\bibinfo  {journal} {Nuclear Physics B}\ }\textbf {\bibinfo {volume} {44}},\ \bibinfo {pages} {189} (\bibinfo {year} {1972})}\BibitemShut {NoStop}%
\bibitem [{\citenamefont {Zinn-Justin}(2021)}]{zinn2021quantum}%
  \BibitemOpen
  \bibfield  {author} {\bibinfo {author} {\bibfnamefont {J.}~\bibnamefont {Zinn-Justin}},\ }\href {https://doi.org/10.1093/oso/9780198834625.001.0001} {\emph {\bibinfo {title} {Quantum Field Theory and Critical Phenomena}}},\ \bibinfo {edition} {5th}\ ed.\ (\bibinfo  {publisher} {Oxford University Press},\ \bibinfo {year} {2021})\BibitemShut {NoStop}%
\bibitem [{\citenamefont {Colizza}\ \emph {et~al.}(2006)\citenamefont {Colizza}, \citenamefont {Barrat}, \citenamefont {Barth{\'e}lemy},\ and\ \citenamefont {Vespignani}}]{colizza2006role}%
  \BibitemOpen
  \bibfield  {author} {\bibinfo {author} {\bibfnamefont {V.}~\bibnamefont {Colizza}}, \bibinfo {author} {\bibfnamefont {A.}~\bibnamefont {Barrat}}, \bibinfo {author} {\bibfnamefont {M.}~\bibnamefont {Barth{\'e}lemy}},\ and\ \bibinfo {author} {\bibfnamefont {A.}~\bibnamefont {Vespignani}},\ }\href {https://doi.org/10.1073/pnas.0510525103} {\bibfield  {journal} {\bibinfo  {journal} {Proceedings of the National Academy of Sciences}\ }\textbf {\bibinfo {volume} {103}},\ \bibinfo {pages} {2015} (\bibinfo {year} {2006})}\BibitemShut {NoStop}%
\bibitem [{\citenamefont {Pastor-Satorras}\ \emph {et~al.}(2015)\citenamefont {Pastor-Satorras}, \citenamefont {Castellano}, \citenamefont {Van~Mieghem},\ and\ \citenamefont {Vespignani}}]{pastorsatorras2015epidemic}%
  \BibitemOpen
  \bibfield  {author} {\bibinfo {author} {\bibfnamefont {R.}~\bibnamefont {Pastor-Satorras}}, \bibinfo {author} {\bibfnamefont {C.}~\bibnamefont {Castellano}}, \bibinfo {author} {\bibfnamefont {P.}~\bibnamefont {Van~Mieghem}},\ and\ \bibinfo {author} {\bibfnamefont {A.}~\bibnamefont {Vespignani}},\ }\href {https://doi.org/10.1103/RevModPhys.87.925} {\bibfield  {journal} {\bibinfo  {journal} {Reviews of Modern Physics}\ }\textbf {\bibinfo {volume} {87}},\ \bibinfo {pages} {925} (\bibinfo {year} {2015})}\BibitemShut {NoStop}%
\bibitem [{\citenamefont {Laskin}(2000)}]{laskin2000fractional}%
  \BibitemOpen
  \bibfield  {author} {\bibinfo {author} {\bibfnamefont {N.}~\bibnamefont {Laskin}},\ }\href {https://doi.org/10.1016/S0375-9601(00)00201-2} {\bibfield  {journal} {\bibinfo  {journal} {Physics Letters A}\ }\textbf {\bibinfo {volume} {268}},\ \bibinfo {pages} {298} (\bibinfo {year} {2000})}\BibitemShut {NoStop}%
\bibitem [{\citenamefont {Olver}\ \emph {et~al.}(2010)\citenamefont {Olver}, \citenamefont {Lozier}, \citenamefont {Boisvert},\ and\ \citenamefont {Clark}}]{olver2010nist}%
  \BibitemOpen
  \bibfield  {author} {\bibinfo {author} {\bibfnamefont {F.~W.~J.}\ \bibnamefont {Olver}}, \bibinfo {author} {\bibfnamefont {D.~W.}\ \bibnamefont {Lozier}}, \bibinfo {author} {\bibfnamefont {R.~F.}\ \bibnamefont {Boisvert}},\ and\ \bibinfo {author} {\bibfnamefont {C.~W.}\ \bibnamefont {Clark}},\ }\href {https://dlmf.nist.gov/} {\emph {\bibinfo {title} {NIST Handbook of Mathematical Functions}}}\ (\bibinfo  {publisher} {Cambridge University Press},\ \bibinfo {address} {New York, NY},\ \bibinfo {year} {2010})\BibitemShut {NoStop}%
\bibitem [{\citenamefont {Bochner}(1949)}]{bochner1949diffusion}%
  \BibitemOpen
  \bibfield  {author} {\bibinfo {author} {\bibfnamefont {S.}~\bibnamefont {Bochner}},\ }\href {https://doi.org/10.1073/pnas.35.7.368} {\bibfield  {journal} {\bibinfo  {journal} {Proceedings of the National Academy of Sciences}\ }\textbf {\bibinfo {volume} {35}},\ \bibinfo {pages} {368} (\bibinfo {year} {1949})}\BibitemShut {NoStop}%
\bibitem [{\citenamefont {Podlubny}(1998)}]{podlubny1998fractional}%
  \BibitemOpen
  \bibfield  {author} {\bibinfo {author} {\bibfnamefont {I.}~\bibnamefont {Podlubny}},\ }\href {https://doi.org/10.1016/S0076-5392(99)X8001-5} {\emph {\bibinfo {title} {Fractional Differential Equations: An Introduction to Fractional Derivatives, Fractional Differential Equations, to Methods of Their Solution and Some of Their Applications}}}\ (\bibinfo  {publisher} {Academic Press},\ \bibinfo {year} {1998})\BibitemShut {NoStop}%
\bibitem [{\citenamefont {Angstmann}\ \emph {et~al.}(2016)\citenamefont {Angstmann}, \citenamefont {Henry},\ and\ \citenamefont {McGann}}]{angstmann2016fractional}%
  \BibitemOpen
  \bibfield  {author} {\bibinfo {author} {\bibfnamefont {C.}~\bibnamefont {Angstmann}}, \bibinfo {author} {\bibfnamefont {B.}~\bibnamefont {Henry}},\ and\ \bibinfo {author} {\bibfnamefont {A.}~\bibnamefont {McGann}},\ }\href {https://doi.org/10.1016/j.physa.2016.02.029} {\bibfield  {journal} {\bibinfo  {journal} {Physica A: Statistical Mechanics and its Applications}\ }\textbf {\bibinfo {volume} {452}},\ \bibinfo {pages} {86} (\bibinfo {year} {2016})}\BibitemShut {NoStop}%
\bibitem [{\citenamefont {Evans}(2010)}]{evans2010partial}%
  \BibitemOpen
  \bibfield  {author} {\bibinfo {author} {\bibfnamefont {L.~C.}\ \bibnamefont {Evans}},\ }\href {https://doi.org/10.1090/gsm/019} {\emph {\bibinfo {title} {Partial Differential Equations}}},\ \bibinfo {edition} {2nd}\ ed.,\ \bibinfo {series} {Graduate Studies in Mathematics}, Vol.~\bibinfo {volume} {19}\ (\bibinfo  {publisher} {American Mathematical Society},\ \bibinfo {year} {2010})\BibitemShut {NoStop}%
\bibitem [{\citenamefont {Kilbas}\ \emph {et~al.}(2006)\citenamefont {Kilbas}, \citenamefont {Srivastava},\ and\ \citenamefont {Trujillo}}]{kilbas2006theory}%
  \BibitemOpen
  \bibfield  {author} {\bibinfo {author} {\bibfnamefont {A.~A.}\ \bibnamefont {Kilbas}}, \bibinfo {author} {\bibfnamefont {H.~M.}\ \bibnamefont {Srivastava}},\ and\ \bibinfo {author} {\bibfnamefont {J.~J.}\ \bibnamefont {Trujillo}},\ }\href {https://doi.org/10.1016/S0304-0208(06)80001-0} {\emph {\bibinfo {title} {Theory and Applications of Fractional Differential Equations}}},\ \bibinfo {series} {North-Holland Mathematics Studies}, Vol.\ \bibinfo {volume} {204}\ (\bibinfo  {publisher} {Elsevier Science B.V.},\ \bibinfo {year} {2006})\BibitemShut {NoStop}%
\bibitem [{\citenamefont {Saeedian}\ \emph {et~al.}(2017)\citenamefont {Saeedian}, \citenamefont {Khalighi}, \citenamefont {Azimi-Tafreshi}, \citenamefont {Jafari},\ and\ \citenamefont {Ausloos}}]{saeedian2017memory}%
  \BibitemOpen
  \bibfield  {author} {\bibinfo {author} {\bibfnamefont {M.}~\bibnamefont {Saeedian}}, \bibinfo {author} {\bibfnamefont {M.}~\bibnamefont {Khalighi}}, \bibinfo {author} {\bibfnamefont {N.}~\bibnamefont {Azimi-Tafreshi}}, \bibinfo {author} {\bibfnamefont {G.~R.}\ \bibnamefont {Jafari}},\ and\ \bibinfo {author} {\bibfnamefont {M.}~\bibnamefont {Ausloos}},\ }\href {https://doi.org/10.1103/PhysRevE.95.022409} {\bibfield  {journal} {\bibinfo  {journal} {Physical Review E}\ }\textbf {\bibinfo {volume} {95}},\ \bibinfo {pages} {022409} (\bibinfo {year} {2017})}\BibitemShut {NoStop}%
\bibitem [{\citenamefont {Lischke}\ \emph {et~al.}(2020)\citenamefont {Lischke}, \citenamefont {Pang}, \citenamefont {Gulian}, \citenamefont {Song}, \citenamefont {Glusa}, \citenamefont {Zheng}, \citenamefont {Mao}, \citenamefont {Cai}, \citenamefont {Meerschaert}, \citenamefont {Ainsworth},\ and\ \citenamefont {Karniadakis}}]{lischke2020what}%
  \BibitemOpen
  \bibfield  {author} {\bibinfo {author} {\bibfnamefont {A.}~\bibnamefont {Lischke}}, \bibinfo {author} {\bibfnamefont {G.}~\bibnamefont {Pang}}, \bibinfo {author} {\bibfnamefont {M.}~\bibnamefont {Gulian}}, \bibinfo {author} {\bibfnamefont {F.}~\bibnamefont {Song}}, \bibinfo {author} {\bibfnamefont {C.}~\bibnamefont {Glusa}}, \bibinfo {author} {\bibfnamefont {X.}~\bibnamefont {Zheng}}, \bibinfo {author} {\bibfnamefont {Z.}~\bibnamefont {Mao}}, \bibinfo {author} {\bibfnamefont {W.}~\bibnamefont {Cai}}, \bibinfo {author} {\bibfnamefont {M.~M.}\ \bibnamefont {Meerschaert}}, \bibinfo {author} {\bibfnamefont {M.}~\bibnamefont {Ainsworth}},\ and\ \bibinfo {author} {\bibfnamefont {G.~E.}\ \bibnamefont {Karniadakis}},\ }\href {https://doi.org/10.1016/j.jcp.2019.109009} {\bibfield  {journal} {\bibinfo  {journal} {Journal of Computational Physics}\ }\textbf {\bibinfo {volume} {404}},\ \bibinfo {pages} {109009} (\bibinfo {year} {2020})}\BibitemShut {NoStop}%
\bibitem [{\citenamefont {Pozrikidis}(2016)}]{pozrikidis2016fractional}%
  \BibitemOpen
  \bibfield  {author} {\bibinfo {author} {\bibfnamefont {C.}~\bibnamefont {Pozrikidis}},\ }\href {https://doi.org/10.1201/b19666} {\emph {\bibinfo {title} {The Fractional Laplacian}}}\ (\bibinfo  {publisher} {CRC Press},\ \bibinfo {year} {2016})\BibitemShut {NoStop}%
\bibitem [{\citenamefont {Diekmann}\ \emph {et~al.}(1990)\citenamefont {Diekmann}, \citenamefont {Heesterbeek},\ and\ \citenamefont {Metz}}]{diekmann1990definition}%
  \BibitemOpen
  \bibfield  {author} {\bibinfo {author} {\bibfnamefont {O.}~\bibnamefont {Diekmann}}, \bibinfo {author} {\bibfnamefont {J.~A.~P.}\ \bibnamefont {Heesterbeek}},\ and\ \bibinfo {author} {\bibfnamefont {J.~A.~J.}\ \bibnamefont {Metz}},\ }\href {https://doi.org/10.1007/BF00178324} {\bibfield  {journal} {\bibinfo  {journal} {Journal of Mathematical Biology}\ }\textbf {\bibinfo {volume} {28}},\ \bibinfo {pages} {365} (\bibinfo {year} {1990})}\BibitemShut {NoStop}%
\bibitem [{\citenamefont {Chaikin}\ and\ \citenamefont {Lubensky}(1995)}]{chaikin1995principles}%
  \BibitemOpen
  \bibfield  {author} {\bibinfo {author} {\bibfnamefont {P.~M.}\ \bibnamefont {Chaikin}}\ and\ \bibinfo {author} {\bibfnamefont {T.~C.}\ \bibnamefont {Lubensky}},\ }\href {https://doi.org/10.1017/CBO9780511813467} {\emph {\bibinfo {title} {Principles of Condensed Matter Physics}}}\ (\bibinfo  {publisher} {Cambridge University Press},\ \bibinfo {year} {1995})\BibitemShut {NoStop}%
\bibitem [{\citenamefont {Binney}\ \emph {et~al.}(1992)\citenamefont {Binney}, \citenamefont {Dowrick}, \citenamefont {Fisher},\ and\ \citenamefont {Newman}}]{binney1992theory}%
  \BibitemOpen
  \bibfield  {author} {\bibinfo {author} {\bibfnamefont {J.~J.}\ \bibnamefont {Binney}}, \bibinfo {author} {\bibfnamefont {N.~J.}\ \bibnamefont {Dowrick}}, \bibinfo {author} {\bibfnamefont {A.~J.}\ \bibnamefont {Fisher}},\ and\ \bibinfo {author} {\bibfnamefont {M.~E.~J.}\ \bibnamefont {Newman}},\ }\href {https://doi.org/10.1093/oso/9780198513940.001.0001} {\emph {\bibinfo {title} {The Theory of Critical Phenomena: An Introduction to the Renormalization Group}}}\ (\bibinfo  {publisher} {Oxford University Press},\ \bibinfo {year} {1992})\BibitemShut {NoStop}%
\bibitem [{\citenamefont {Hufnagel}\ \emph {et~al.}(2004)\citenamefont {Hufnagel}, \citenamefont {Brockmann},\ and\ \citenamefont {Geisel}}]{hufnagel2004forecast}%
  \BibitemOpen
  \bibfield  {author} {\bibinfo {author} {\bibfnamefont {L.}~\bibnamefont {Hufnagel}}, \bibinfo {author} {\bibfnamefont {D.}~\bibnamefont {Brockmann}},\ and\ \bibinfo {author} {\bibfnamefont {T.}~\bibnamefont {Geisel}},\ }\href {https://doi.org/10.1073/pnas.0308344101} {\bibfield  {journal} {\bibinfo  {journal} {Proceedings of the National Academy of Sciences}\ }\textbf {\bibinfo {volume} {101}},\ \bibinfo {pages} {15124} (\bibinfo {year} {2004})}\BibitemShut {NoStop}%
\bibitem [{\citenamefont {Goldenfeld}(1992)}]{goldenfeld1992lectures}%
  \BibitemOpen
  \bibfield  {author} {\bibinfo {author} {\bibfnamefont {N.}~\bibnamefont {Goldenfeld}},\ }\href {https://doi.org/10.1201/9780429493492} {\emph {\bibinfo {title} {Lectures on Phase Transitions and the Renormalization Group}}}\ (\bibinfo  {publisher} {CRC Press},\ \bibinfo {year} {1992})\BibitemShut {NoStop}%
\bibitem [{\citenamefont {Song}\ \emph {et~al.}(2005)\citenamefont {Song}, \citenamefont {Havlin},\ and\ \citenamefont {Makse}}]{song2005self}%
  \BibitemOpen
  \bibfield  {author} {\bibinfo {author} {\bibfnamefont {C.}~\bibnamefont {Song}}, \bibinfo {author} {\bibfnamefont {S.}~\bibnamefont {Havlin}},\ and\ \bibinfo {author} {\bibfnamefont {H.~A.}\ \bibnamefont {Makse}},\ }\href {https://doi.org/10.1038/nature03248} {\bibfield  {journal} {\bibinfo  {journal} {Nature}\ }\textbf {\bibinfo {volume} {433}},\ \bibinfo {pages} {392} (\bibinfo {year} {2005})}\BibitemShut {NoStop}%
\bibitem [{\citenamefont {Althouse}\ \emph {et~al.}(2020)\citenamefont {Althouse}, \citenamefont {Wenger}, \citenamefont {Miller}, \citenamefont {Scarpino}, \citenamefont {Allard}, \citenamefont {H{\'e}bert-Dufresne}, \citenamefont {Hu} \emph {et~al.}}]{althouse2020superspreading}%
  \BibitemOpen
  \bibfield  {author} {\bibinfo {author} {\bibfnamefont {B.~M.}\ \bibnamefont {Althouse}}, \bibinfo {author} {\bibfnamefont {E.~A.}\ \bibnamefont {Wenger}}, \bibinfo {author} {\bibfnamefont {J.~C.}\ \bibnamefont {Miller}}, \bibinfo {author} {\bibfnamefont {S.~V.}\ \bibnamefont {Scarpino}}, \bibinfo {author} {\bibfnamefont {A.}~\bibnamefont {Allard}}, \bibinfo {author} {\bibfnamefont {L.}~\bibnamefont {H{\'e}bert-Dufresne}}, \bibinfo {author} {\bibfnamefont {H.}~\bibnamefont {Hu}}, \emph {et~al.},\ }\href {https://doi.org/10.1371/journal.pbio.3000897} {\bibfield  {journal} {\bibinfo  {journal} {PLOS Biology}\ }\textbf {\bibinfo {volume} {18}},\ \bibinfo {pages} {e3000897} (\bibinfo {year} {2020})}\BibitemShut {NoStop}%
\bibitem [{\citenamefont {Lloyd}(2001)}]{lloyd2001realistic}%
  \BibitemOpen
  \bibfield  {author} {\bibinfo {author} {\bibfnamefont {A.~L.}\ \bibnamefont {Lloyd}},\ }\href {https://doi.org/10.1006/tpbi.2001.1525} {\bibfield  {journal} {\bibinfo  {journal} {Theoretical Population Biology}\ }\textbf {\bibinfo {volume} {60}},\ \bibinfo {pages} {59} (\bibinfo {year} {2001})}\BibitemShut {NoStop}%
\bibitem [{\citenamefont {Morawska}\ and\ \citenamefont {Milton}(2020)}]{morawska2020time}%
  \BibitemOpen
  \bibfield  {author} {\bibinfo {author} {\bibfnamefont {L.}~\bibnamefont {Morawska}}\ and\ \bibinfo {author} {\bibfnamefont {D.~K.}\ \bibnamefont {Milton}},\ }\href {https://doi.org/10.1093/cid/ciaa939} {\bibfield  {journal} {\bibinfo  {journal} {Clinical Infectious Diseases}\ }\textbf {\bibinfo {volume} {71}},\ \bibinfo {pages} {2311} (\bibinfo {year} {2020})}\BibitemShut {NoStop}%
\bibitem [{\citenamefont {Barab{\'a}si}(2005)}]{barabasi2005origin}%
  \BibitemOpen
  \bibfield  {author} {\bibinfo {author} {\bibfnamefont {A.-L.}\ \bibnamefont {Barab{\'a}si}},\ }\href {https://doi.org/10.1038/nature03459} {\bibfield  {journal} {\bibinfo  {journal} {Nature}\ }\textbf {\bibinfo {volume} {435}},\ \bibinfo {pages} {207} (\bibinfo {year} {2005})}\BibitemShut {NoStop}%
\bibitem [{\citenamefont {Riley}(2007)}]{riley2007large}%
  \BibitemOpen
  \bibfield  {author} {\bibinfo {author} {\bibfnamefont {S.}~\bibnamefont {Riley}},\ }\href {https://doi.org/10.1126/science.1134695} {\bibfield  {journal} {\bibinfo  {journal} {Science}\ }\textbf {\bibinfo {volume} {316}},\ \bibinfo {pages} {1298} (\bibinfo {year} {2007})}\BibitemShut {NoStop}%
\bibitem [{\citenamefont {Glass}\ \emph {et~al.}(2006)\citenamefont {Glass}, \citenamefont {Glass}, \citenamefont {Beyeler},\ and\ \citenamefont {Min}}]{glass2006targeted}%
  \BibitemOpen
  \bibfield  {author} {\bibinfo {author} {\bibfnamefont {R.~J.}\ \bibnamefont {Glass}}, \bibinfo {author} {\bibfnamefont {L.~M.}\ \bibnamefont {Glass}}, \bibinfo {author} {\bibfnamefont {W.~E.}\ \bibnamefont {Beyeler}},\ and\ \bibinfo {author} {\bibfnamefont {H.~J.}\ \bibnamefont {Min}},\ }\href {https://doi.org/10.3201/eid1211.060255} {\bibfield  {journal} {\bibinfo  {journal} {Emerging Infectious Diseases}\ }\textbf {\bibinfo {volume} {12}},\ \bibinfo {pages} {1671} (\bibinfo {year} {2006})}\BibitemShut {NoStop}%
\bibitem [{\citenamefont {Eubank}\ \emph {et~al.}(2004)\citenamefont {Eubank}, \citenamefont {Guclu}, \citenamefont {Kumar}, \citenamefont {Marathe}, \citenamefont {Srinivasan}, \citenamefont {Toroczkai},\ and\ \citenamefont {Wang}}]{eubank2004modelling}%
  \BibitemOpen
  \bibfield  {author} {\bibinfo {author} {\bibfnamefont {S.}~\bibnamefont {Eubank}}, \bibinfo {author} {\bibfnamefont {H.}~\bibnamefont {Guclu}}, \bibinfo {author} {\bibfnamefont {V.~A.}\ \bibnamefont {Kumar}}, \bibinfo {author} {\bibfnamefont {M.~V.}\ \bibnamefont {Marathe}}, \bibinfo {author} {\bibfnamefont {A.}~\bibnamefont {Srinivasan}}, \bibinfo {author} {\bibfnamefont {Z.}~\bibnamefont {Toroczkai}},\ and\ \bibinfo {author} {\bibfnamefont {N.}~\bibnamefont {Wang}},\ }\href {https://doi.org/10.1038/nature02541} {\bibfield  {journal} {\bibinfo  {journal} {Nature}\ }\textbf {\bibinfo {volume} {429}},\ \bibinfo {pages} {180} (\bibinfo {year} {2004})}\BibitemShut {NoStop}%
\end{thebibliography}
%\include{bib}

%apsrev4-2.bst 2019-01-14 (MD) hand-edited version of apsrev4-1.bst
%Control: key (0)
%Control: author (72) initials jnrlst
%Control: editor formatted (1) identically to author
%Control: production of article title (-1) disabled
%Control: page (0) single
%Control: year (1) truncated
%Control: production of eprint (0) enabled
%
\end{document}